\documentclass{article}
\usepackage[utf8]{inputenc}
\usepackage{amsmath}
\usepackage{amsthm}
\usepackage{amssymb}
\usepackage{hyperref}
\usepackage{nameref}
\usepackage{pdfsync}
\usepackage{mathrsfs}
\usepackage{textcomp}
\usepackage{xcolor}
\usepackage{authblk}
\usepackage{graphicx}
\usepackage[normalem]{ulem}
\usepackage[colorinlistoftodos]{todonotes}
\usepackage{tcolorbox}

\newcount\Comments  % 0 suppresses notes to selves in text
\usepackage{color}
\usepackage{titlesec}
\usepackage{diagbox}
\definecolor{darkgreen}{rgb}{0,0.5,0}
\definecolor{purple}{rgb}{1,0,1}
\definecolor{teal}{rgb}{0,0.4627,0.5804}
\usepackage{xspace}
\usepackage{lipsum}
\usepackage{graphicx}
\usepackage{etoolbox}
\makeatletter
\providecommand{\subtitle}[1]{% add subtitle to \maketitle
  \apptocmd{\@title}{\par {\large #1 \par}}{}{}
}
\makeatother

\newcommand{\kibitz}[2]{\ifnum\Comments=1\textcolor{#1}{#2}\fi}

\begin{document}

\newtcolorbox{mybox}{colback=red!5!white,colframe=red!75!black}
\newtcolorbox{pullquote}{colback=blue!5!white,colframe=blue!75!black}

\date{\empty}

\title{Traffic smoothing using explicit local controllers}
\subtitle{Dissipating Stop-and-Go Waves with a Single Automated Vehicle in Dense Traffic: Experimental Evidence}

\author{AMAURY HAYAT\textsuperscript{\textdagger}, ARWA ALANQARY\textsuperscript{*,1}, 
RAHUL BHADANI\textsuperscript{$\S$, 1}, CHRISTOPHER DENARO\textsuperscript{\textdaggerdbl\textdaggerdbl,1}, RYAN J. WEIGHTMAN\textsuperscript{\textdaggerdbl\textdaggerdbl,1}, SHENGQUAN XIANG\textsuperscript{\textbardbl,1}, 
JONATHAN W.~LEE    \textsuperscript{*}, 
    MATTHEW BUNTING\textsuperscript{\textdaggerdbl},
    ANISH GOLLAKOTA\textsuperscript{*},
    MATTHEW W.~NICE\textsuperscript{\textdaggerdbl},
    DEREK GLOUDEMANS\textsuperscript{\textdaggerdbl},
    GERGELY ZACH\'{A}R\textsuperscript{\textdaggerdbl},
    JON F. DAVIS\textsuperscript{*},
    MARIA LAURA DELLE MONACHE\textsuperscript{*},
    BENJAMIN SEIBOLD\textsuperscript{**},
    ALEXANDRE M.~BAYEN\textsuperscript{*},
    JONATHAN SPRINKLE\textsuperscript{\textdaggerdbl},
    DANIEL B.~WORK\textsuperscript{\textdaggerdbl}
    AND
    BENEDETTO PICCOLI\textsuperscript{\textdaggerdbl\textdaggerdbl},
%Benjamin Seibold, Maria Laura Delle Monache, Dan Works, Jonathan Sprinkle, Alexandre M. Bayen, Benedetto Piccoli
}

\affil{
    \textsuperscript{*}--University of California, Berkeley\\
    \textsuperscript{\textdagger}--\'Ecole des Ponts Paristech, Marne la Vall\'ee\\
    \textsuperscript{\textdaggerdbl}--Vanderbilt University\\ 
    \
    \textsuperscript{\textbardbl}-- Peking University\\
    \textsuperscript{$\S$} -- The University of Alabama in Huntsville\\
%    \textsuperscript{\P}--Queen's University\\
    \textsuperscript{**}--Temple University\\  \textsuperscript{\textdaggerdbl\textdaggerdbl}--Rutgers University-Camden\\
    \textsuperscript{1}--These authors contributed equally\\
}

% MATH Macros for consistency and if we change them later

\maketitle

%\dois{}{}

%\textcolor{red}{\chapterinitial{T}he abstract goes there}\emph{} 

\begin{abstract}
%\textcolor{red}{\summaryinitial{T}he Summary sidebar must be be the first item cited in the text, before any other sidebars, figures, or tables. It is a brief description of the article.}
The dissipation of stop-and-go waves attracted recent attention as a traffic management problem, which can be efficiently addressed by automated driving. As part of the 100 automated vehicles experiment named MegaVanderTest, feedback controls were used to induce strong dissipation via velocity smoothing. More precisely, a single vehicle driving differently in one of the four lanes of I-24 in the Nashville area was able to regularize the velocity profile by reducing oscillations in time and velocity differences among vehicles. Quantitative measures of this effect were possible due to the innovative I-24 MOTION~\cite{gloudemans202324} system capable of monitoring the traffic conditions for all vehicles on the roadway. 
%The reconstructed vehicle trajectories exhibit the reduced speed oscillation behavior behind the autonomous vehicle compared with those in front.
This paper presents the control design, the technological aspects involved in its deployment, and, finally, the results achieved by the experiment.
\end{abstract}

\section{Introduction}
%\textcolor{red}{Todo}\\
%\jms{@amaury: can you confirm the authors list?} \ah{here is an attempt, there might be one or two additional authors. Of course feel free to tell me if you think there is anyone missing.}
%\textcolor{red}{Stop-and-go wave, stakes and interest of dissipating them}

Stop-and-go waves are ubiquitous traffic instabilities observed in almost all parts of the world
\cite{treiterer1974hysteresis, kerner2012physics}. The drawbacks of such waves include increased fuel consumption and decreased safety \cite{STERN2018205}. Taming or dissipating them is a  problem of traffic management. The latter has witnessed a revolution with the technical capability of replacing fixed control actuators (for example, toll gates, traffic lights, and traffic signals) with mobile actuators
such as connected and automated vehicles (AVs). 
Researchers studied the capability of this control paradigm to regulate traffic, including dissipating waves, even if most results are for the case of high penetration of AVs (high percentage of AVs as part of the bulk traffic) or completely autonomous traffic.
Simulation results can be found in 
\cite{PhysRevE.69.066110,TALEBPOUR2016143,GUERIAU2016266,7362183}.
Other strategies included 
variable speed limit strategies,
\cite{BDSH08,WDHA16,HAN2017113} and
jam absorption
\cite{NISHI2013116,7534847}.
The modeling of the problem generates some interesting mathematical challenges, such as the need for multiscale models
\cite{GGLP20}.

In the last few years, some experimental results have become available
\cite{STERN2018205,WU201982,STERN2019351}. 
In particular, the ring-road
experiment described in these papers shows how a single AV can tame waves produced by 21 human-driven vehicles.
The experiment setup replicated faithfully the seminal one described in \cite{sugiyama2008traffic}:
22 vehicles started at equal distances on a circular track of 260 meters and reached the same speed. After a short time, stop-and-go waves naturally appear but the single AV is able to dissipate them. These results were achieved originally by the controller described in \cite{bhadani2018dissipation}. 
% JMS: the original version of this made it sound as if the following citations contained controllers used in the ring road experiment
Continued research has explored either
% via various controllers designed either by 
model-based techniques \cite{albeaik2022limitations,delle2019feedback,giammarino2020traffic,Hayat2023theory,Kardous2022multi} or AI ones \cite{lichtle2022deploying,vinitsky2018benchmarks,wu2017emergent,wu2017flowOld,wu2021flow,yan2022unified}.
The encouraging results included a decrease in speed variance, reduced fuel consumption, and reduced heavy braking. In this experiment, drivers were instructed to close the gap while driving safely but with no knowledge of the experiment's goals. This reduced the potential biases but limitations remained: the confined setting, the single-lane situation, and the artificial environment.

In order to move forward to an open highway, a holistic vision was proposed \cite{Hayat2022holistic,lee2021integrated} based on an innovative monitoring camera system \cite{gloudemans202324},
the use of advanced hardware devices \cite{bunting2021libpanda,bhadani2022strym,nice2023middleware},
energy modeling and control algorithms.
The approach consisted of inserting 100 automated vehicles in bulk traffic and using various control algorithms. The controller presented in this paper was used on a single vehicle out of the 100.
%\textcolor{red}{
%Overall 100 AVs
Since the expected penetration rate was around 1-3\%, each AV was surrounded by 30-100 human-driven vehicles. With this in mind, the key problem was that of designing controls for a single AV in a multilane setting (with no lane-changing maneuvers), which would impact traffic while 
% keeping 
maintaining requirements for 
safety. As for the ring-road experiments, the focus was on smoothing traffic via reduced speed variance. Also, in this setting, the smoothed traffic was expected to be more fuel-efficient and safe than oscillating ones and stop-and-go waves.

% \textcolor{red}{As part of this approach message Single  vehicle.}

%\section{Contribution of this paper}
In this paper, we present the experimental evidence that a single automated car equipped with an appropriate model-based controller can efficiently dissipate 
%TODO be more specific / say more than that
a stop-and-go wave in open traffic on the highway. This experiment was carried out in November 2022 as part of the MegaVanderTest that took place on I-24 in Nashville, TN, described further in \cite{AmeliLivetrafficCSM}. The highway includes freight traffic, exhibits stop-and-go traffic daily during rush hour, and includes more than 150,000 vehicles daily. 
This is the ideal situation for testing the controls and the general idea of traffic smoothing via a small number of AVs. 

% we will undoubtedly cite this paper, but it doesn't make as much sense as the scenarios paper.
% \cite{gloudemans202324}.

We first present the controller we used, which was designed from mathematical principles using a microscopic traffic model \cite{Hayat2023theory}. To smooth traffic while ensuring safety, the controller was designed as a combination of three parts: a safety module, a target speed, and a Model Predictive Control (briefly MPC) component. The safety module computes the maximal speed, which would allow the AV to avoid collision in case of sudden braking of the leading vehicle (directly in front of the AV). The target speed is the expected uniform speed the smoothed traffic will travel at. Such speed is either decided using global information (speed planner) or local traffic velocity. Finally, the Model Predictive Control component is designed to anticipate the speed changes of the leader vehicle while keeping the speed as close as possible to the target ones.

This paper goes beyond the theoretical development of the controller. It demonstrates implementation on a full-sized car, deployed in traffic as a moving traffic wave controller. 
%\jms{I removed some text here that we cover. At this point I think just delays the onset of the paper's interesting content, with marginal benefit to the reader.}
% theoretically designed controller needed to be implemented on a real car, specifically a Toyota Rav 4. The following measurements were available: GPS coordinates and instantaneous velocity of the car, as well as the distance to the leading vehicle. Such measurements are typically used by the Adaptive Cruise Control (ACC) system of the Rav 4.
% First, the control was coded using Simulink and with input/output using ROS (Robot Operating System) and validated using the Gazebo-based simulator \cite{Bhadani_2018}. Then the hardware implementation was based on Libpanda \cite{bunting2021libpanda} and CAN\_to\_ROS interface \cite{elmadani2021can}.
The experiment was conducted as part of the MegaVanderTest. The deployment was on the westbound I-24 highway to Nashville, TN, on Wednesday, November 16th, during morning rush hour (8-9 am), on a 9.33-kilometer stretch between exits 66 and 60. 
% A Toyota Rav 4 was human-driven from the headquarters, located near exit 60, to the I-24. After reaching the assigned lane (not an HOV one) the driver was activating the control via the standard ACC button. The driver was instructed to keep the control engaged as long as it was deemed safe. 
We analyze the results using the trajectories reconstructed by the I-24 MOTION system \cite{gloudemans202324}. Each car trajectory was obtained through processing and analysis of video data from the myriad cameras of I-24 MOTION (see \hyperlink{sidebar:I-24sidebar}{I-24 MOTION}).

Since our aim was traffic smoothing via speed oscillation reduction, we focused on computing the speed variance along trajectories. More specifically, the effect of the AV control is measured by comparing the speed variance of the cars running in front of the AV (thus not subject to the control) with those of the cars running behind the AV. A direct effect can be visually noted by comparing the lane where the AV traveled with other lanes (see Figure \ref{fig:time_space_new}). 

We demonstrate our results by calculating and comparing speed variance. The speed variance over 1.4 km behind the AV was 50\% less than the speed variance in front. A more detailed analysis, see Table \ref{tab:metrics_sv}, reveals that such impact is stronger in the vicinity of the AV (200 to 400 m). Indeed oscillations are observed to appear again around 600 m behind the AV. This length has to be compared with the ring-road experiment of \cite{STERN2018205,WU201982,STERN2019351}, where oscillations were completely removed on a length of 260 m. In simple words, the AV obtained a similar effect on an open highway with a strong effect for around double the size of the ring road.

In the remainder of this paper, we will carefully review the design, implementation, deployment, and analysis of our experiment and its results. The paper will show how a carefully crafted model-based acceleration controller was able to smooth traffic on an open highway with dense traffic and four lanes. The main result is the cut in half the speed oscillations between vehicles in front and behind the AV. Moreover, the controller used standard sensor measurements from stock ACC systems on commercially available Toyota Rav4 models, thus opening the door to large-scale implementation.\\

%\textcolor{red}{Introduce and coordinate what is next.}

% \textcolor{red}{
% Box on control, principle of what we are doing.}

% \begin{pullquote}
% We design and implement an anticipation controller
% which enables a single automated vehicle to dissipate a stop-and-go traffic wave and significantly reduce speed oscillations on the highway.
% \end{pullquote}

% \begin{pullquote}
% We present experimental evidence that an AV running a model-based controller can locally dissipate a stop-and-go wave in heavy traffic on the highway.
% %We present the experimental evidence that a single automated car equipped with an appropriate model-based controller can efficiently dissipate a stop-and-go wave in open traffic on the highway.
% \end{pullquote}

\begin{mybox}
\section*{Control theory, stabilization... and road traffic\label{side:control}}
\noindent
\emph{by Amaury Hayat and Shengquan Xiang}

\subsection*{Control theory in a nutshell...}
%[Control and traffic]
Control theory is about asking: ``If I can act on the system, what can I make it do?". From a mathematical point of view, it consists of having a system
\begin{equation}
    \dot x(t)  = f(x(t),u(t)),
\end{equation}
where $x(t)$ is the state and $u(t)$ is a function --called control-- that can be chosen and represent the way we can act on the system. A typical goal in control theory is to know, given an initial state $x_{0}$, what states $x_{1}$ can be reached by choosing $u(t)$ properly.

Stabilization is a sub-branch of control theory where the goal is to make sure that the system follows a target state $\bar{x}$ and returns to it when disturbed. A control is designed to stabilize a system that would be unstable in the absence of a controlling force.   That is to say
\begin{gather}
\left\{\begin{split}
&\text{for any }\varepsilon>0,\text{ there exists }\eta\text{ s.t. for all }t\in[0,+\infty)\\
&\|x(0)-\bar{x}(0)\|\leq \eta\; \implies\; \|x(t)-\bar{x}(t)\|\leq \varepsilon,
\end{split}
\right.
\\
\exists \delta>0\text{ s.t. }\|x(0)-\bar{x}(0)\|\leq \delta\;\implies \;\lim\limits_{t\rightarrow+\infty}\|x(t)-\bar{x}(t)\| = 0.
\end{gather}

Most of the time, $\bar{x}$ is chosen as a constant and is an equilibrium of the system, that is 
\begin{equation}
    f(\bar{x},0) = 0.
\end{equation}
The particularity of stabilization is that the control $u(t)$ does not depend on the initial condition $x_{0}$ but rather on the current state $x(t)$, that is formally
\begin{equation}
u(t) = g(x(t)).
\end{equation}
In more general versions $u(t)$ could also depend on past state $(x(\tau))_{\tau\leq t}$. A control in this form is called a \emph{feedback law}.
\subsection*{...and in road traffic}
In road traffic, using an AV to smooth stop-and-go waves enters the following framework: $x(t)$ represents the state of the cars on the road, for instance, their position $h_{i}$ and velocity $v_{i}$. The control $u(t)$ is the acceleration or the velocity of the AV that can be chosen, up to some safety and hardware constraints. Formally, this can be written as
\begin{equation}
    \underbrace{\dot x(t)}_{\text{traffic state}} = \underbrace{f(x(t), \underbrace{u(t)}_{\text{AV dynamic}})}_{\text{traffic dynamic}}.
\end{equation}
\end{mybox}
\begin{mybox}
The target state $\bar{x}$ is the uniform flow equilibrium where every car is going the same constant speed. This target state is unstable in congestion when there is no controlled AV, meaning that $x(t)$ is usually far from its equilibrium value $\bar{x}$. An example of such a system in traffic is, for instance (see \cite{bando1995dynamical,gazis1961nonlinear,Hayat2023theory,Gong2023})
\begin{equation}
\left\{\begin{split}
\dot v_{0}& = u(t),\\
\dot y_{0}&= v_{0},\\
\dot v_{i}&=a\frac{v_{i+1}-v_{i}}{(y_{i+1}-y_{i})^{2}}+b[V(y_{i+1}-y_{i})-v_{i}],\\
\dot y_{i}&=v_{i},\;\;\;\;\
\label{sys01}
\end{split}\right.
\end{equation}

Smoothing stop-and-go waves simply amounts to choosing $u$ to reduce %\todo[author=Xiang]{It would look slightly more fancy if we replace the following by $\int_{0}^T \|x(t)- \bar x\| dt$?}
$$
\int_{0}^T \|x(t)-\bar{x}\|dt,
$$
where the state is $x(t) = (y(t),v(t))$ and $T$ is a given time horizon.
In practice, there are several difficulties: 
\begin{itemize}
\item[(i)] The target state $\bar{x}$ (and in particular the target speed) is usually unknown in practical cases such as a highway. Two ways to tackle this difficulty can be thought of:
\begin{itemize}
\item Infer a good approximation of $\bar{x}$ from both theory and experimental measurements. This is, in part, the principle of the speed planner presented below (see \hyperlink{side:planner}{Hierarchical Control Framework and Speed Planner}); 
\item Aim to reduce the variance of $x$ with respect to time instead of aiming to reduce $\|x(t)-\bar{x}\|$. 
%This has the advantage of not needing 
%Assuming that the only steady-state is $\bar{x}$ this 
\end{itemize}
\item[(ii)] The dynamic $f$ is quite complicated because of the lane changes: the system is usually either \emph{infinite dimensional} or \emph{hybrid}.
\item[(iii)] The mathematical models representing road traffic are usually imprecise. Thus the control needs to be robust with respect to errors on $f$.
\item[(iv)] The control system must exhibit robustness in handling errors related to measurements, including signal loss, sensor and camera limitations, etc. 
\end{itemize}
\end{mybox}

%\clearpage

\section{Controller design}
From a control theory perspective, the approach consists of considering a connected AV as a means of control on the system, that is, the whole traffic flow. 
%This is described below in \nameref{side:control}. 
The controller is acceleration-based, meaning that the control variable is the acceleration of the connected AV.
In this section, we describe the AV's acceleration-based controller based on the design described in \cite{Microaccelth}.
%, and designed specifically to smooth stop-and-go waves.

% \renewcommand{\thesequation}{S\arabic{equation}}
% % \setcounter{stable}{0}
% \renewcommand{\thestable}{S\arabic{stable}}
% % \setcounter{sfigure}{0}
% \renewcommand{\thesfigure}{S\arabic{sfigure}}

\subsection{Principle}
\label{sec:principle}
The controller combines three components:

\begin{itemize}
\item \textbf{Safety}, a module that ensures that the vehicle never puts itself or others in danger.
\item \textbf{Target}, a module that calculates the target speed required to achieve the control goal.
\item  \textbf{Model Predictive Control (briefly MPC)}, a module that anticipates the leader's behavior to help limit the AV's speed deviations from the target speed.
\end{itemize}

Each of the modules leads to a limit acceleration respectively denoted by $a_{\text{safe}}$, $a_{\text{target}}$, $a_{\text{MPC}}$. The controller combines these three limit accelerations by taking their minimum.
%More precisely 
The commanded acceleration of the controller can be written at each time as:
\begin{equation}
    a_{\text{cmd}}(t) = \min(a_{\text{safe}}(t),a_{\text{target}}(t),a_{\text{MPC}}(t)),
\end{equation} 
%where $a_{\text{safe}}$ is the acceleration guaranteeing safety,  ($i.e. \; a_{\text{safe}}$), $a_{\text{target}}$ is the ideal desired control acceleration to reach the control objective, and $a_{\text{MPC}}$ is the MPC  based acceleration which takes into account the anticipation of the leader's behavior. 
The mathematical expression of $a_{\text{safe}}(t)$, $a_{\text{target}}(t)$ and $a_{\text{MPC}}(t)$ are detailed in the following paragraphs.

%\textbf{Notations}
\subsection{Notations}
We introduce the following notations for these time-varying signals: 
\begin{itemize}
\item $v$ refers to the instantaneous driving speed of the AV (or ego vehicle).
\item $v_{\text{lead}}$ is the measured speed of the leader vehicle.
%\item $x_{\text{rel}}$ is the space gap distance between ego vehicle and its leader.
\item $h$ is the space gap between the ego vehicle and the leading vehicle.
\item $v_{\text{rel}}$ is the relative speed of the leader with respect to the ego vehicle.
\item $a$ is the measured estimate of acceleration of the ego vehicle. 
\item $a_{\text{lead}}$ is the actual acceleration of the leading vehicle.
\end{itemize}

%\ah{Introduce the notations $v$, $a$, etc.}
%First of all, we care about the safety of our system. This is a local model since if there is no cut in whether the ego vehicle will crush in  the leading vehicle  depends only on the  local information, namely the current speed and acceleration 
\subsection{Safety Module}
We first define $v_{\text{safe}}$ as the highest velocity below which the AV can remain safe by braking if needed, whatever the behavior of the leading vehicle. Under this velocity, even if the leader brakes extremely strongly until full stop, the AV can avoid collision. 

This $v_{\text{safe}}(t)$ is a computed value at any time $t$, 
depending on the space gap $h(t)$, the maximal braking capacity of the AV (a constant, denoted $a_{\min}<0$), and the maximal braking capacity of the leading vehicle (a constant denoted $a_{l,\min}<0$). This velocity can be computed explicitly and is given by (see \cite[Theorem 1.1]{Microaccelth})
\begin{equation}
\label{eq:vsafe}
v_{\text{safe}}(t)= \sqrt{ 2|a_{\min}|\left(h(t)-s_{0}+\frac{1}{2} \frac{v_{\text{lead}}^{2}(t)}{|a_{l,\min}|}\right)},
\end{equation}
where $s_{0}>0$ is a given safety distance.
The acceleration $a_{\text{safe}}(t)$ is then defined as 
\begin{equation}
\label{eq:asafe}
a_{\text{safe}}(t) = -k(v(t)-v_{\text{safe}}(t))+\frac{dv_{\text{safe}}(t)}{dt},
\end{equation}
where $k \in \mathbb{R}^{+}$ is a given positive parameter. This acceleration acts as a barrier and guarantees the safety of the AV (provided that it starts in the safe area and no unsafe lane changes happen, see \cite[Theorem 1.1]{Microaccelth}).
% and $v_{\text{safe}}$ is the highest velocity that guarantees the safety of $AV$, whatever the behavior of the leading vehicle (see \cite[Theorem 1.1]{}Microaccelth for a theoretical proof). This is given by 
% \begin{equation}
% \label{eq:vsafe}
% v_{\text{safe}}(t)= \sqrt{ 2|a_{\min}|\left(h(t)-s_{0}+\frac{1}{2} \frac{v_{lead}^{2}(t)}{|a_{l,\min}|}\right)}.
% \end{equation}

\subsection{Target Module}
The acceleration $a_{\text{target}}(t)$ is defined as 
\begin{equation}
a_{\text{target}}(t) = -k(v(t)-v_{\text{target}}(t)),
\end{equation}
where $v_{\text{target}}(t)$ is a target velocity chosen at time $t$ to reach the control goal. The choice of this target velocity depends on the availability of the downstream information. We consider two modes:

\textbf{Local mode}
When there is no downstream information available, either because there is no source or because of a defect in the connectivity of the AV, the target speed is chosen from the only information available, that is, the velocity of the leading vehicle, the one of the AV, and the space gap. In this case, the goal is to use this information to reconstruct an approximation of what would be the steady-state speed if there were as many vehicles but no stop-and-go wave. This speed is chosen as follows
%\jms{@amaury, why $\bar v(\cdot)$ here, vs. just $v(\cdot)$?}
\begin{equation}
v_{target}(t) =  \bar{v}_{\text{lead}}(t)
+ c_{1}\frac{\max(0,c_{2}(h(t)-\delta_{1}v(t)))}{\max(1,v(t)))^{2}},
\end{equation}
where 
\begin{equation}
\label{eq:bvlead}
 \bar{v}_{lead}(t)=   
 \left\{ 
 \begin{split}
 \frac{1}{t}\int_{0}^{t} v_{\text{lead}}(s) ds, \text{ if } t\leq \tau, \\
  \frac{1}{\tau}\int_{t-\tau}^{t}v_{\text{lead}}(s) ds, \text{ if } t> \tau,
 \end{split}
 \right.
\end{equation}
and $c_{1}$, $\delta_{1}$ and $\tau$ are design parameters that can be chosen and have the following interpretation:
\begin{itemize}
    \item $c_{1}$ is a catching-up weight,
    \item $\delta_{1}$ is a target time gap between the ego and the leading car,
    \item $\tau$ is an approximation of the period of a stop-and-go wave.
\end{itemize}

%\textcolor{red}{XXX}
%\arwa{Define constants and functions: $c_2, h(t) \delta_1, \tau$}

\textbf{Planning mode} 
When downstream information is available, 
%more precisely $v_{\text{traffic}}(t,x)$ is the estimated speed of the flow at location $x$ and time $t$, 
we use the speed planner described in \hyperlink{side:planner}{Hierarchical Control Framework and Speed Planner}
%\jms{@amaury are these other CSM papers?} \ah{Yes ! We will add them soon}
to get an estimation of the target speed for the current location of the ego vehicle based on the downstream traffic information. We denote this speed by $v_{\text{down}}(t)$, and we use it as our target speed after clipping it to be within some factor of the leader speed and ensure that it does not exceed a reference upper-bound speed $v_{\text{ref}}$. That is
\begin{equation}
\begin{split}
\bar{v}_{\text{target}}(t)&= \max(\textcolor{black}{\max}(v_{\text{down}}, \textcolor{black}{\alpha_{0}}v_{\text{lead}}(t)), \min(\textcolor{black}{\alpha_{1}} v_{lead}(t),v_{ref})),
%v_{down}&=\min_{x\in[0,L]}\left(\int_{\tau_{0}}^{t-\tau_{0}}v_{\text{traffic}}(s,x)ds\right)
\end{split}
\end{equation}
where  $\alpha_{0}\in(0,1)$ and $\alpha_{1}>1$ are design parameters and $v_{\text{ref}}$ is a reference velocity which corresponds to a safeguard when we know an upper bound of the traffic speed in congestion. If no such bound is known, $v_{\text{ref}}$ is simply set to the road speed limit.
%\ah{@Shengquan and @Arwa, are we sure of that ? Weren't we were using directly the output of the speed planner in $v_{\text{down}}$. 
%Add link between $v_{\text{??}}$
%}
%\arwa{%Yes we did use the speed planner output for $v_{donw}$. 
%I'm wondering if the speed planner sidebar should include the logic for how this is being calculated because it currently doesn't.
%Also give some context for the physical meaning of $v_{ref}$
%}
%\ah{I agree.}
%\jms{@arway or @amaury, can you resolve this as you see fit by the end of the weekend ending on 9/24?}
%\ah{Should be good now.}

\subsection{MPC Control}
The MPC control is the most complex part. It is designed to anticipate the leader's behavior and restrict the AV's deviation from its target speed. The paradigm is: to react quickly to a change in the leader's behavior, but as little and smoothly as possible. %with the 

To do so, when the leading vehicle decelerates, the acceleration $a_{MPC}$ commanded by the MPC module is set as the smallest possible deceleration such that the AV will remain safe in terms of collision: in other words, it would not reach the safety distance, should the leader keep its constant deceleration until full stop.

%and is designed as follows:
%\textcolor{red}{explain the logic}\\
To compute this value, 
%do so, when the leading vehicle decelerates, 
we define the acceleration $a_{\text{min brake}}$ as: 
%\jms{What is the meaning of the `12' subscript? can we call it `safe' or something?} \ah{I changed the name, I'll try to figure a better one, safe would collide with $a_{\text{safe}}$}
\begin{equation}
\label{eq:a12}
\begin{split}
&a_{\text{min brake}}(h(t),v(t),v_{\text{lead}}(t),a_{\text{lead}}(t)) \\
&=-\left(h(t)-s_{0}+\frac{1}{2} \frac{v_{\text{lead}}^{2}(t)}{-a_{\text{lead}}(t)}\right)^{-1} \frac{(v(t))^{2}}{2},
\end{split}
\end{equation}
and the quantities
\begin{align}
P_1=& \; a_{\text{min brake}}(h(t),v(t),v_{\text{lead}}(t),a_{\text{lead}}(t))\\
& \; -a_{\text{lead}}(t)v(t)/v_{\text{lead}}(t), \nonumber\\
P_2=& \; v_{\text{lead}}(t)- v(t).
\end{align}
The acceleration $a_{MPC}$ commanded by the anticipation module is
\begin{equation}
 a_{MPC}= \left\{
\begin{split}
 &a_{\text{min brake}}(h(t),v(t),v_{\text{lead}}(t),a_{\text{lead}}(t)), \text{ if } P_1>0, \\
&a_{\text{lead}}v(t)/v_{\text{lead}}(t), \text{ if } P_1\leq 0 \text{ and } P_2\geq 0, \\
&a_{\text{lead}}-\frac{(v-v_{\text{lead}})^{2}}{2(h(t)-s_{0})}, \text{ if } P_1\leq 0 \text{ and } P_2< 0. 
\end{split}
 \right.
\end{equation}
   
 When the leading vehicle speeds up, to avoid any unwanted behavior and jittering we ensure the continuity of the controller by setting
\begin{equation}
 a_{MPC}= \left\{
\begin{split}
&a_{\text{lead}}-\frac{(v-v_{\text{lead}})^{2}}{2(h(t)-s_{0})}, \text{ if }  P_2< 0, \\
&\min(a_{\max}, a_{\text{lead}}(1+k_{2}(v_{\text{lead}}-v))), \text{ if }  P_2\geq  0. 
\end{split}
 \right.
\end{equation}

%\textcolor{red}{Explain why this complicated formula} 
More details and a detailed theoretical analysis of this controller can be found in \cite{Microaccelth}.\\

\begin{mybox}
\section*{Hierarchical Control Framework and Speed Planner \hypertarget{side:planner}{}}
\noindent
\emph{by Han Wang}
%\vspace{-5mm}
%\section[the Speed planner Sidebar]{}
%\subsection*{}

% \setcounter{equation}{0}

%\begin{figure}
{\centering
\includegraphics[width=19.0pc]{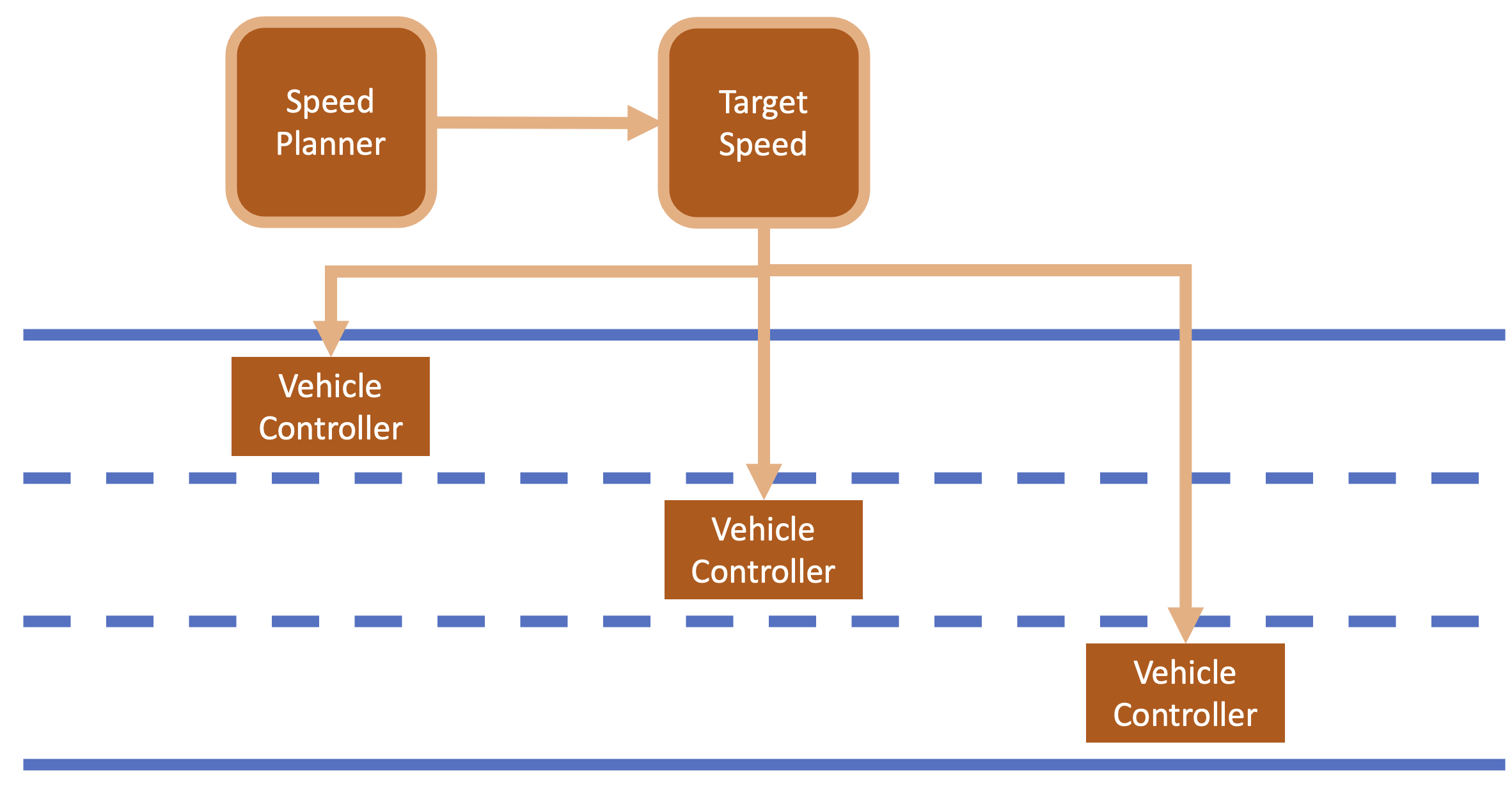}\par}

\textbf{Centralized Speed Planner:} the algorithm deployed on the server side that designs target speed profiles for vehicles in moving traffic, with the goal to smooth traffic waves. The system was validated on a fleet of 100 connected automated vehicles in the MVT\\
%\end{figure}

The proposed controller served as the vehicle side controller in a novel hierarchical control framework. The framework includes two critical components: 1) the collection of algorithms on the server side operates as the centralized planner agent dealing with the heavy calculation tasks; 2) the algorithms deployed on the vehicle side act as executors following the target assigned by the centralized planner. The principle of the framework design is to solve the computational task allocation problem between the server and the onboard units (OBU) and to efficiently coordinate the control goals between macroscopic traffic flow optimization and microscopic vehicle control. 

At the beginning of each update, the speed planner extracts a combination of macroscopic traffic state estimation(TSE) and vehicle observations from the database to calculate the target speed profile. According to the demand and condition of the specific implementation, the speed profile could be published in various formats. Each vehicle's OBU will fetch the most recent target speed from the target speed profile and use it as the input, together with the local observation from the onboard detector's perceptions. 

For the Buffer Design module, we consider the interval $\mathcal I \subset \mathbb{R}$ as the region of interest. Suppose further that $\mathcal I_c \subset \mathcal I$ denotes a congested area. The idea is to determine the moving bottleneck (the controlled vehicle) speed profile denoted by $t \mapsto U_b(t)$, such that the density $k(t, x)$ for $x \in \mathcal I_c$ is distributed (evenly) throughout the entire region $\mathcal I$ and consequently by average the density approaches $k_c$, the critical density associated with maximum flow. Determining the moving bottleneck speed profile will be done in the following steps: (i) predicting the density $(t, x) \mapsto k(t,x)$, $(t, x) \in \mathbb{R}_+ \times \mathcal I$ given a speed profile $U_b(\cdot)$ of the controlled vehicle, using a mathematical model of traffic flow (see the next subsection), (ii) assessing the efficiency of the speed profile $U_b(\cdot)$ based on the density $k(t,x)$ employing the reinforcement learning (RL), and (iii) Updating $U_b(\cdot)$ and returning to step (i).

The vehicle controllers extract information from the target speed and local observation for the control action selection. The local observation will be submitted to the central database as the future input for the speed planner.
\end{mybox}

\subsection{Adaptation to loss of signal}

%\paragraph{Robustness to Lack of Signal}
Between the ideal mathematical framework and reality, there are many disturbances and unplanned constraints. To be able to work in real life, the control deployed has to be robust to a number of external perturbations. As underlined in \cite{Microaccelth}, this controller is robust to delay in the measurements or the actuation or to small measurement imprecisions. However, as it stands, this controller is not robust to the loss of signal from the front sensor, which could lead to highly overestimating the space gap and the control to overshoot by far. This loss of signal could be the consequence of a road grade, a curve, or simply a radar malfunction or range limitations. 
An instance of this was encountered several times during the experiment. 
%Due to hardware limitations, a loss of signal from the front sensor would occur if the space gap to the car in front was greater than some (unknown) limit.

%\textcolor{red}{To mitigate the effect of this on the performance of the controller, when such loss of signal is detected, the }\\

Over and above this problem, the question arises as to what control strategy to adopt when the position of the leading car is unknown for a long period. A simple option would consist of assuming that the leading car has a space gap and velocity that are identical to the last space gap and velocity measured until the signal is reached again. That is to say
\begin{equation}
v_{\text{lead}}(t) = v_{\text{lead}}(t_{1}),\;\;\; h(t) = h(t_{1}),\;\text{ for any } t\in[t_{1},t_{2}),
\end{equation}
where $t_{1}$ is the time of the last signal measured and $t_{2}$ is the time of the next signal measured.
This limits jumps in the behavior of the controlled AV.

However, in \textit{local mode} (see Section \textit{Principle}), this velocity $v_{\text{lead}}(t)$ is used to estimate the ideal steady-state speed of the system. Assuming a constant velocity equal to the last measured velocity of the leading car could bias this estimation by having a strong weight in \eqref{eq:bvlead}. This could lead to underestimating the real flow speed of the traffic and entering a self-perpetuating situation where the automated vehicle is slower than the traffic and consequently keeps too large a gap to the car in front for the radar to notice. To tackle this issue, for each time step where the signal is lost, we gradually increase the last seen space gap by setting 
\begin{align}
    h(t+\Delta t) = h(t) + \Delta t(h_{correction}). 
\end{align}
%\jms{I changed this from h(t+1) to h(t + dt). Please confirm} AH: I confirm, thanks !
This results in increasing the estimated $v_{safe}$ and $\bar{v}_{target}$, which in turn allows the ego vehicle to gradually increase its velocity and close the gap to its leader. 

In the \textit{planning mode}, we clip the target velocity to be no more than $\alpha_1 v_{lead}$. So if the leader speed is severely underestimated (due to the lost signal), it can lead to an underestimation of the target velocity. To overcome this, we remove this upper bound on the target speed when the signal is lost and allow it to be larger than the last leader's speed observed until the signal is regained.

\begin{pullquote}
{\Large
``The controller that combines safety, objectives, and anticipation is capable of operating both with and without downstream information and remains robust even in the event of a signal loss."
\par}

\end{pullquote}

\section{Software and hardware implementation}

%\textcolor{red}{TODO Rahul}

\begin{figure}[t]
\centering
\includegraphics[width=0.99\linewidth, trim={0cm 0.0cm 0cm 0.0cm},clip]{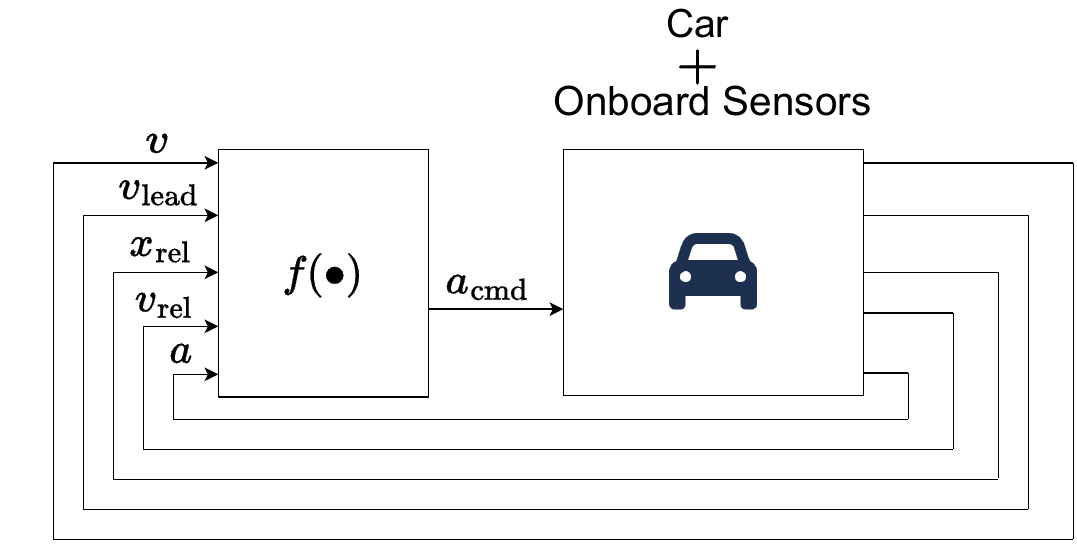}
\caption{Dataflow layout of the controller in \eqref{eq:energycontroller}, applied to traffic control. The controller is designed for the longitudinal movement of the ego vehicle to follow its leader, which may be a human-driven vehicle.}
\label{fig:Microaccel_block_diagram}
\end{figure}

We use a model-based design approach to implement the controller discussed earlier. We can abstract the controller as 
\begin{equation}
\label{eq:energycontroller}
a_{\text{cmd}}(t) = f( v(t), v_{\text{lead}}(t), h(t), v_{\text{rel}}(t), a(t); \Theta).
\end{equation}
Recall that $v$ is the instantaneous driving speed of the vehicle to be controlled (ego vehicle), $v_{\text{lead}}$ is the measured speed of the leader vehicle, $h(t)$ is the space gap between ego vehicle and its leader, $v_{\text{rel}}$ is the relative speed of the leader with respect to the ego vehicle, and $a$ is the acceleration of the ego vehicle. In \eqref{eq:energycontroller}, the controller parameters are represented as $\Theta$. The functional layout of the controller model is given in Figure~\ref{fig:Microaccel_block_diagram}.

\subsection{Simulink with Code Generation to ROS}
The abstracted controller \eqref{eq:energycontroller} is implemented as a Simulink model with data input and output components modeled using ROS (Robot Operating System)~\cite{quigley2009ros} Toolbox. The Simulink model is used to generate a standalone C++ ROS node that can be executed directly on a physical hardware board without any modification. Moving from the Simulink model to a C++ ROS node without writing any C++ code allows for faster prototyping and validation at an early stage in the simulation through data-driven software-in-the-loop validation of the controller behavior. The ROS node consists of multiple ROS subscribers that consume input data and provide acceleration command output through a ROS publisher~\cite{bhadani2023approaches}. %\rahul{I added most stuff I wanted except two things: (i) frequency of messages (ii) all parameters used for the controller. For (i), I want to verify from bagfiles what's the frequency of each topic but i couldn't get the hold of the bag file; (ii) for parameter values, I will leave Amaury to cross-reference from the function, otherwise if I need to specify parameters then I need to still talk to Amaury to make sure the variables in the MATLAB function matches the notation used in the paper which I am not sure of.}
ROS nodes are executed in a parameterized manner through \texttt{roslaunch}, 
% \texttt{roslaunch} is an XML 
a configuration-based tool for ROS that
% to allow 
allows the execution of multiple ROS with node parameters supplied at runtime.
% of ROS nodes are supplied during runtime. 
% For the controller execution, \texttt{roslaunch} executes the controller and speed planner with a single configuration file. 
Validation testing used regression data played back through our Gazebo-based simulator \cite{Bhadani_2018}, where 
% We tested the controller using  \texttt{roslaunch} in tandem with the Gazebo simulator~\cite{koenig2004design} that let us create a 
multi-vehicle simulation uses rigid body dynamics. This allowed us to compare the software deployment candidate with desired performance criteria prior to deploying in hardware.

\begin{mybox}

\section*{I-24 MOTION \hypertarget{sidebar:I-24sidebar}{}}

\noindent
\emph{by Derek Gloudemans, Gergely Zach\'{a}r, Yanbing Wang, Junyi Ji, Will Barbour, Dan Work}\\
%\section[I-24-Motion Sidebar]{}

I-24 MOTION~\cite{gloudemans202324,gloudemans2020interstate} is a 4.2-mile instrumented section of I-24 in Nashville, TN. It serves as a traffic data collection instrument and a test bed for connected and automated vehicle technologies and traffic control strategies. The instrumentation consists of 276 4K resolution video cameras mounted on 110-135ft roadside poles. The cameras provide a complete view of the roadway and can observe the path of all vehicles with minimal occlusion. 

A video processing pipeline consisting of computer vision~\cite{gloudemans2023so,gloudemans2023interstate,gloudemans2021vehicle} and post-processing algorithms identifies and tracks vehicle locations, then stitches~\cite{wang2023onlinemcf} and reconciles their trajectories to ensure physical feasibility. The trajectory reconciliation consists of optimization-based smoothing that ensures feasible vehicle dynamics in higher orders~\cite{wang2022automatic}. Trajectories are generated in a roadway-aligned curvilinear coordinate system \cite{gloudemans2023so} that can be converted to or from GNSS coordinate systems for the purposes of aligning data sources measured from vehicles. From the I-24 MOTION data, multiple traveling traffic waves are regularly observed during morning peak commute times, leading to high-speed variability and consequently increased fuel consumption.\\

% \sdbarfig{\includegraphics[width=19.0pc]{figures/placeholder_I24.png}}{A time-space diagrams of traffic trajectories reconstructed from I24-MOTION. The color indicate the velocity at a given point.\label{fig:sidebarI24M}}

% \sdbarfig{\includegraphics[width=\linewidth]{figures/MOTION-viz.png}}{Time-space diagram of traffic trajectories reconstructed from I24-MOTION, with the AV trajectory overlaid in light blue. Insets show close-ups of traffic near the AV.}{\label{fig:sidebarI24M}}

%\sdbarfigfullwidth{
%\begin{figure}
    {\centering
\includegraphics[width=0.9\textwidth]{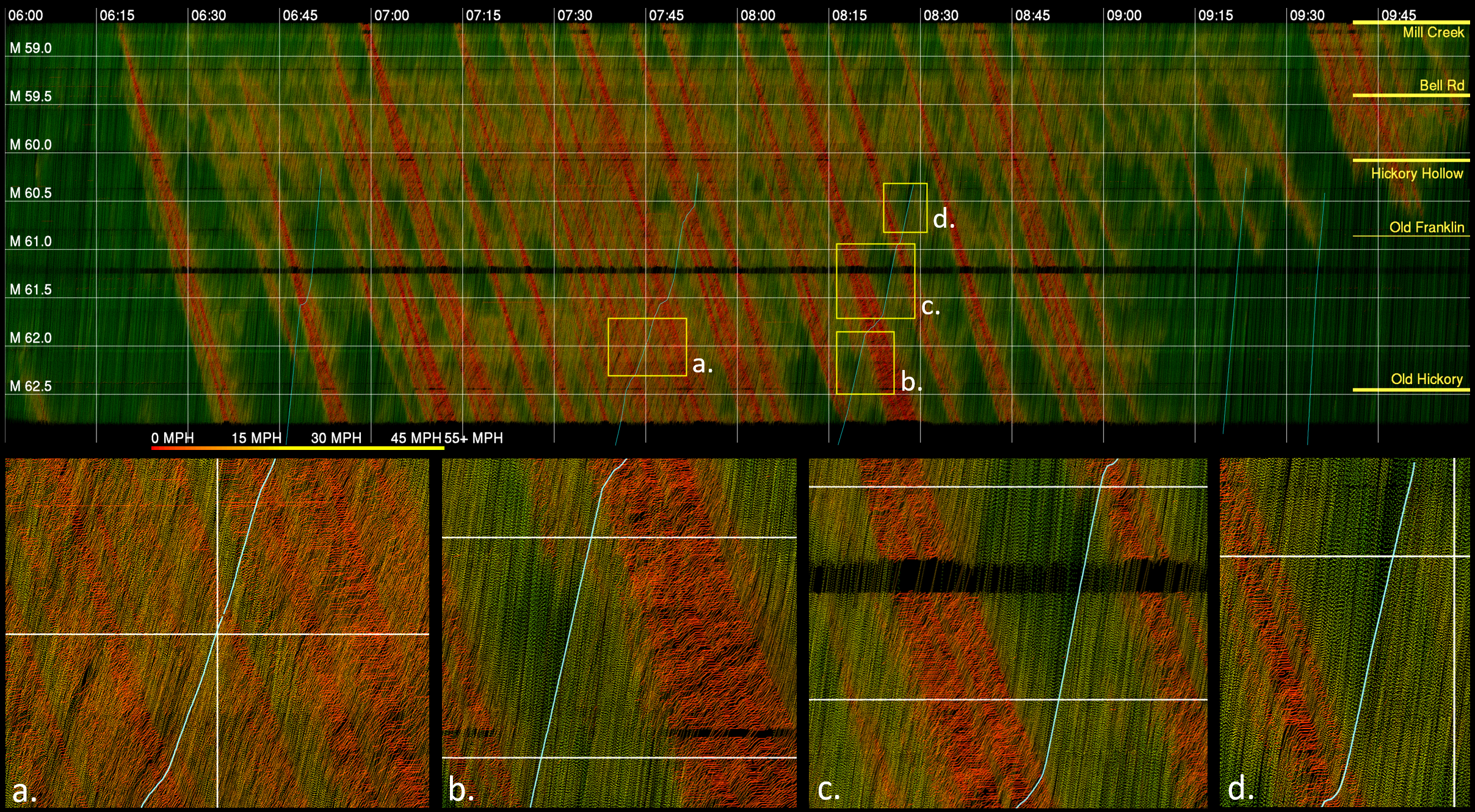}\par}

\textbf{Time-space diagram of traffic trajectories reconstructed from I24-MOTION}~\cite{gloudemans202324} and visualized with~\cite{10.1145/3576914.3587710}, with the AV trajectory overlaid in light blue. Insets show close-ups of traffic near the AV.
    \label{fig:sidebarI24M}\\
    
%\end{figure}

% \sdbarfigfull{\includegraphics[width=\linewidth]{figures/MOTION-viz.png}}{Time-space diagram of traffic trajectories reconstructed from I24-MOTION, with the AV trajectory overlaid in light blue. Insets show close-ups of traffic near the AV.}{\label{fig:sidebarI24M}}

% \begin{figure*}
%     \includegraphics[width=\textwidth]{figures/MOTION-viz.png}
%     \caption{Time-space diagram of traffic trajectories reconstructed from I24-MOTION, with the AV trajectory overlaid in light blue. Insets show close-ups of traffic near the AV. 
%     \label{fig:sidebarI24M}}
% \end{figure*}

\end{mybox}

\subsection{Deployment to Vehicle Platform}
The ego vehicle (the test vehicle on which the controller was deployed) was a Toyota RAV4, capable of acceleration-based control using our customized hardware and software stack. As described further in \cite{LeeMegacontrollerCSM} the hardware stack includes Controller Area Network (CAN) transceivers for access to vehicle data (including on-board sensors and actuators) and to inject vehicle commands. The software stack relies on Libpanda \cite{bunting2021libpanda} to read from hardware. The package strym \cite{bhadani2022strym} was used at runtime to decode the on-board data from the CAN, and the CAN\_to\_ROS package \cite{elmadani2021can,nice2023middleware} was used to transmit data into ROS for producers/consumers of vehicle data. This included the ability to share the live view of the vehicle's state and to allow automatic upgrades, as described in \cite{nice2023middleware}. All of the software and hardware are integrated through a Raspberry Pi, enabling a cost-effective means to retrofit the vehicle for data acquisition and control.
% is equipped with a Raspberry Pi board and a CAN-harness device \textbf{Comma.AI Panda}. Comma.AI Panda device connects the test vehicle's CAN bus peripherals to the Raspberry Pi for vehicle data transfer for logging and making control decisions. The data logging is enabled by a high-performance C++ library Libpanda~\cite{bunting2021libpanda} installed on the Raspberry Pi that reads incoming datastream from the CAN peripherals. Libpanda consists of two observers to
% USB data, used for GPS and CAN data. They can be used for data polling at regular time intervals. The data is consumed by CAN\_to\_ROS package~\cite{elmadani2021can}, also installed on the Raspberry Pi, and publishes data relevant to the controller's decision-making on ROS publishers.

\begin{figure}[h]
\centering
\includegraphics[width=0.99\linewidth, trim={0cm 0.0cm 0cm 0.0cm},clip]{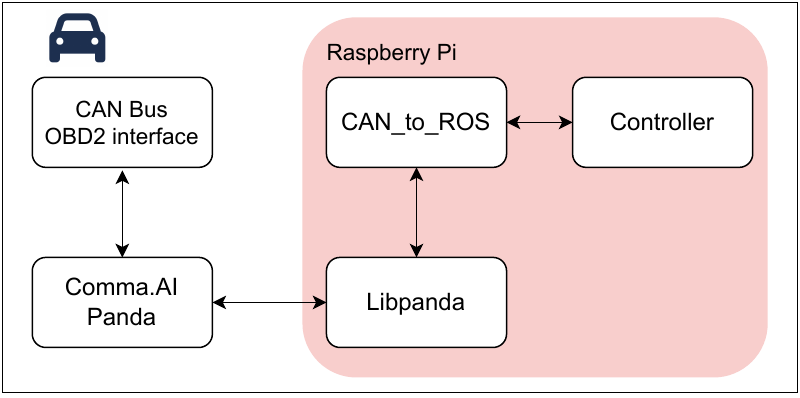}
\caption{A functional layout of the controller for car-following strategy in a mixed-autonomy traffic experiment. The controller is designed for the longitudinal movement of the ego vehicle to follow its leader, which may be a human-driven vehicle.\label{fig:CANDataFlow}}
\end{figure}

The role of Libpanda and CAN\_to\_ROS is illustrated in Figure~\ref{fig:CANDataFlow}. The vehicle interface node in CAN\_to\_ROS subscribes to the commanded acceleration topic, and converts the required command to CAN messages that are sent over to the vehicle via CAN peripherals for actuation. A more detailed description of the controller implementation can be found in~\cite{bhadani2023approaches}.

\section{Framework of the experiment}
\label{sec:framework}
\begin{figure}[h]
\centering
\includegraphics[width=0.99\linewidth, trim={0cm 0.0cm 0cm 0.0cm},clip]{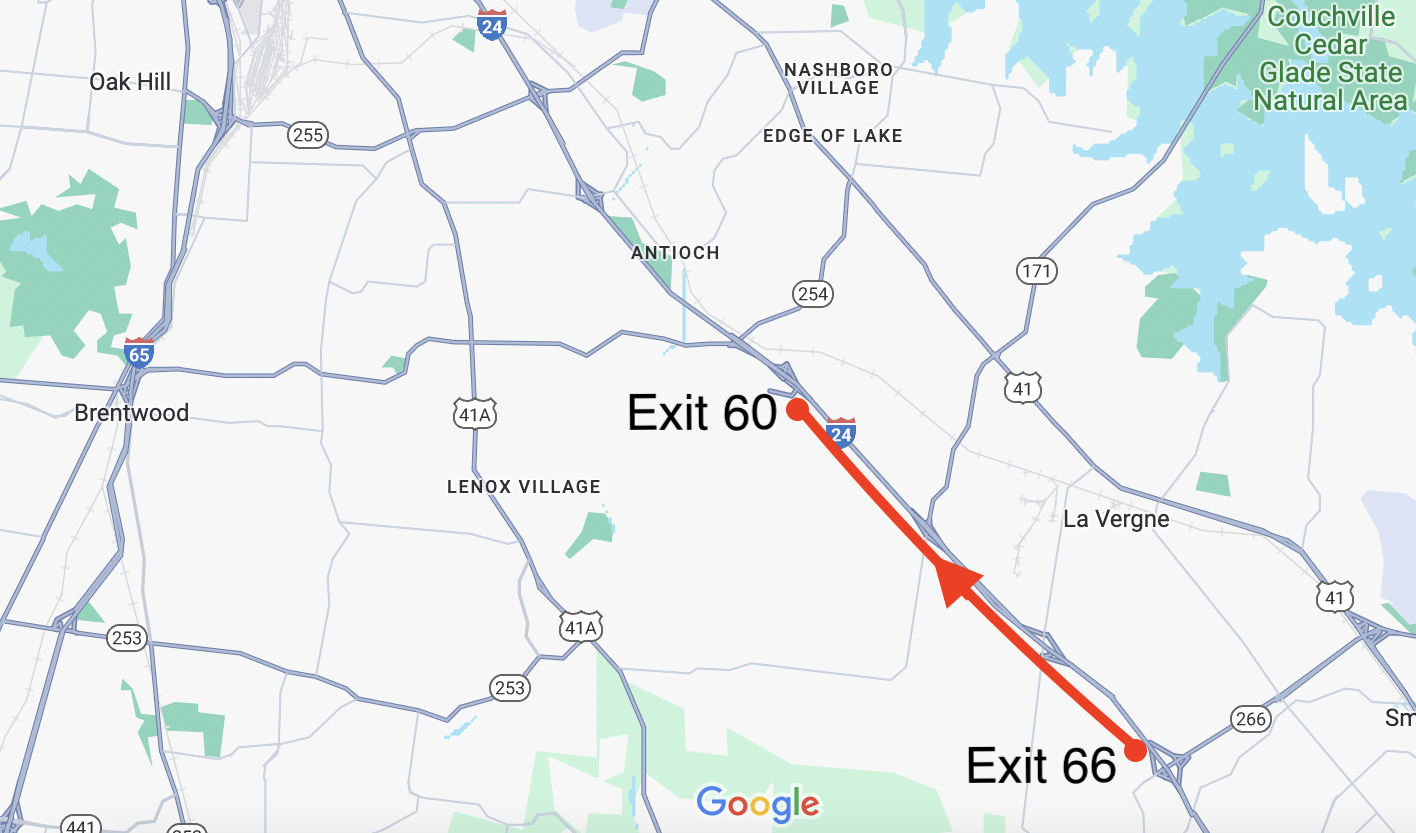}
\caption{The experiment took place on the westbound of the I-24 between exits 66 and 60 with a total distance of approximately 9.33 km. \label{fig:map}}
\end{figure}

The controller was deployed on a Toyota RAV4 in dense traffic on a segment of the four-lane westbound I-24 highway to Nashville, TN (see Figure \ref{fig:map}). The experiment was carried out on Wednesday, November 16th, during the morning rush hour. The results presented in this paper are collected between 08:10 am and 08:50 am.
%were carried out over the span of 4 days (November 14th and 16th-18th 2022). 
%\arwa{Novermber 17, 18, and 19, 2023?}. 
During this time, the AV was traveling westbound between exits 66 and 60 for a total distance of 9.33 km and was located in lane 3, where lane 4 denotes the rightmost lane and lane 1 denotes the leftmost lane.
%the experiments ran during the morning rush hour (between 6-8am) on westbound I-24. 

During the experiment runs, the controller sets the desired acceleration of the vehicle, but only when Adaptive Cruise Control (ACC) is enabled and engaged: thus, the driver interface is the same as they would normally expect. Vehicle design features allow the controller to initially engage once the vehicle is driving above $20 mph$. The driver engaged the controller as soon as the vehicle is driving above this limit on the highway. 
Both in stock ACC and with this controller, some events can trigger the controller to disengage without the driver's input. 
%(due to the reason's mentioned above) 
For instance, a close cut-in in front of the ego vehicle or following a very strong deceleration by the lead vehicle.
%\arwa{Should we mention the issues we had with forced disengagement of the controller: at lane changes and when the vehicle slows down?}\ah{Maybe very briefly as limitations, to explain that in this case we ask the driver to follow the control as most as possible.}
In this case, the driver establishes as soon as possible, while under human control, an appropriate gap that allows for re-engaging the controller. In instances where the average speed of traffic was expected to be less than $20 mph$ for some time due to congestion and the controller disengaged for one of the reasons above, the driver was asked to manually follow the commanded values asked by the controller with the help of the on-board computer monitored by a researcher on the passenger seat until it could be re-engaged.

Data from the experimental runs are collected in two ways. The CAN data from the ego vehicle is logged in a database as in \cite{bunting2022data}.
More importantly, the trajectories of the AV and surrounding vehicles are recovered from video footage recorded and processed by I-24 MOTION~\cite{gloudemans202324} to extract the trajectories. A brief description of the I-24 MOTION system and data set can be found in \hyperlink{sidebar:I-24sidebar}{I24-MOTION} and more details can be found in \cite{gloudemans202324}, in particular concerning the reconstruction procedure of the trajectories. The AV trajectory obtained from the camera is then synchronized with the AV trajectory from the CAN data to avoid any spatial offset due to GPS imprecision.

\begin{pullquote}
{\Large
``We present experimental evidence that an AV running a model-based controller can locally dissipate a stop-and-go wave in heavy traffic on the highway."\par}
%We design and implement a controller that combines safety, objectives, and anticipation. This controller is capable of operating both with and without downstream information and remains robust even in the event of a signal loss.
\end{pullquote}

\begin{figure*}[t]
    \centering
    \includegraphics[width=\linewidth]{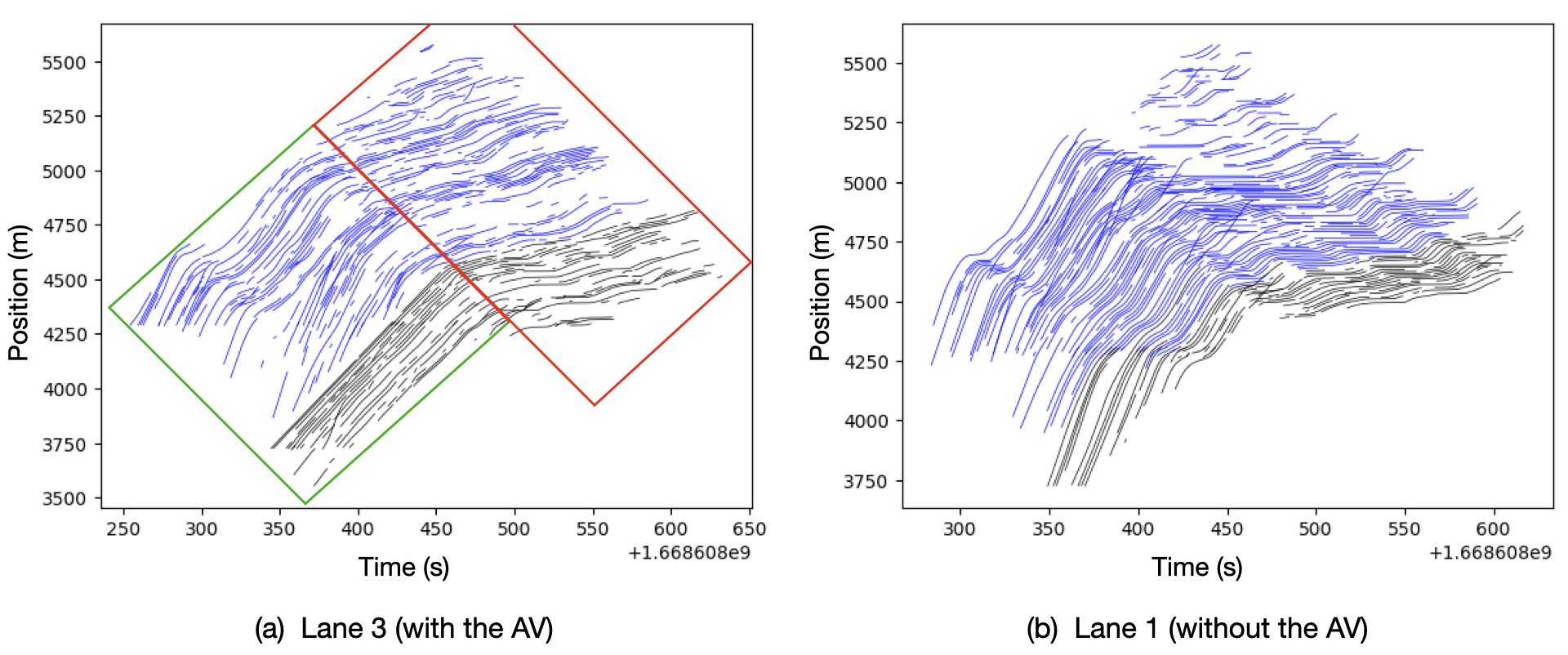}
    \caption{Time-space diagrams showing the trajectories of the vehicles up to 400 m upstream (black) and up to 1400 m downstream (blue) of the AV on two different lanes. \textbf{(a):} The trajectories are shown from the AV's lane (lane 3). \textbf{(b):} The trajectories are shown from another lane without an AV (lane 1). The green %\textcolor{orange}{green} 
    box on figure (a) indicates the region where the AV completely absorbs the wave} 
    \label{fig:time_space_new}
\end{figure*}

%\clearpage

\section{Results}
In this section, we present the experimental results of the controller on the test run described in \ref{sec:framework} section. These results show evidence that the single AV deployed in dense traffic improves the performance of the system locally. We will exhibit its effect qualitatively by looking into the time-space diagrams in the vicinity of the AV. We will also quantify the effect using the speed variance. 

\subsection{Time-space diagrams.}
%%% Describe the AV lane figure %%%
In Figure \ref{fig:time_space_new} (a), we show the trajectories of the AV and the surrounding vehicles in the same lane (lane 3) through a single wave. We include trajectories up to $800 \ m$ upstream of the wave bottleneck. The trajectories in black denote the vehicles behind the AV, while the trajectories in blue are the vehicles in front of the AV. We show trajectories up to $1400\ m $ in front of the AV and $400\ m$ behind it. Because the wave propagates backward in time, we need to identify the moving border of the wave. The procedure to do so is detailed in Appendix \ref{app:cutting}.

%%% Describe the no AV figure%%%
In Figure \ref{fig:time_space_new} (b), we show the trajectories of the vehicles in the same time-space region of the identified wave in a different lane (lane 1), which does not contain any AVs. We use lane 1 for comparison (instead of the adjacent lane 2) to reduce the possible spillover effect of the AV on adjacent lanes. Similarly, in this figure, the black trajectories denote the vehicles upstream of the $x$-position of the AV, while blue trajectories are the downstream vehicles. 

%%% Analyze the AV figure %%%
From Figure \ref{fig:time_space_new} (a), we notice that the blue trajectories in front of the AV experience speed oscillations due to the propagation of waves
similar to Figure \ref{fig:time_space_new} (b). However, the effect of the AV is apparent on the vehicles behind it (black trajectories). In the region %\textcolor{orange}{
upstream the bottleneck (the green box of Figure \ref{fig:time_space_new} (a))
%}
%between $x=4100 - 4500m$, and $t=280 - 350s$, 
%\textcolor{orange}{
the oscillations are
%} 
completely absorbed by the AV and do not propagate to the vehicles behind it. In this region, we notice that the AV is driving with a steady speed despite the stop-and-go behavior in front of it. The trailing vehicles' behavior is closely aligned with the AV's as they also exhibit %\textcolor{orange}{
much
%} 
less speed variations. The AV eventually catches up with the bottleneck (the red region of the time-space diagram) and has to slow down, indicating that it is traveling faster than the bottleneck.

%%% Analyze the no AV figure %%%
In contrast, we notice from Figure \ref{fig:time_space_new} (b) that in the absence of the AV, the wave propagates throughout the 
%\textcolor{orange}{
entire
%} 
region. We observe several stop-and-go oscillations that affect all the vehicles in that region. 

\begin{table*}[ht]
    \centering
    \begin{tabular}{|c|c|c|c|c|c|c|c|}
    \hline \diagbox{Distance behind AV}{Distance in front of AV}
    &1400m & 1200m & 1000 m & 800m & 600m & 400m & 200m \\
    \hline
200m &-58\% & -58\% &-59\% & -62\% & -67\%& -71\% & -74\%\\
\hline
%300m & & & & \%\\
%\hline
400m &-52\% &-52\% & -53\% &-56\%& -62\%& -66\%& -70\% \\
\hline
600m &-52\% &-51\% &-52\% &-56\%& -62\%& -66\% & -69\%\\
\hline
800m &-41\% &-40\% &-41\% & -46\%& -53\%& -59\% & -63\%\\
\hline
1000m &-22\% &-22\% &-23\% & -30\%& -39\%& -46\% & -51\%\\
\hline
1200m &-11\% &-10\% &-12\%\% & -19\%& -30\%& -38\% & -44\%\\
\hline
1400m &+5\% &+6\% &+5\%\% & +4\%& -17\%& -27\% & -34\%\\
\hline
    \end{tabular}
%%%%% VALUES %%%%%
%1400 19.6
%1200 19.4
%1000 19.8 
%800 21.6
%(600 24) Probably too close from the AV and the 200 first meter influence by the AV have a strong effect.

%-200 8.2
%-400 9.4
%-600 9.5
%-800 11.6
%-1000 15.2
%-1200 17.4
%-1400 20.5
%
\vspace{\baselineskip}
    \caption{Percentage change of speed variance of the vehicles behind the controlled AV compared to the vehicles in front of it up to different distances. }
    \label{tab:metrics_sv}
\end{table*}

\subsection{Speed Variance}
To explore the smoothing effect of the AV, we compare the speed variance of the vehicles behind the AV to those in front of the AV. We use the trajectories shown in Figure \ref{fig:time_space_new} (a) to compute the speed variance where the variance is computed across all cars and all time steps (for the vehicles in front and behind the AV separately). 
%
% The speed variance of the trajectories \textbf{in front} of the AV is $\boldsymbol{19.6}\ m^s \cdot s^{-2}$ while the speed variance for the trajectories \textbf{behind} the AV is $\boldsymbol{9.4}\ m^s \cdot s^{-2}$. 
% JMS changed the units
The speed variance of the trajectories \textbf{in front} of the AV is $\boldsymbol{19.6}\ m^2 \cdot s^{-2}$ while the speed variance for the trajectories \textbf{behind} the AV is $\boldsymbol{9.4}\ m^2 \cdot s^{-2}$. 
The introduction of the AV has a smoothing effect that reduces the speed variance by $\boldsymbol{52\%}$. This is aligned with the observed behavior in the time-space diagram is \ref{fig:time_space_new} (a). 

\vspace{-2mm}
\subsection{Discussion}
The results presented in the previous section clearly demonstrate that the proposed controller has a substantial impact on dissipating stop-and-go waves and reducing speed variance. 
%To further reinforce these results we consider the 
%\textcolor{orange}{
Even in the most bottleneck part of the wave (the red region in Figure \ref{fig:time_space_new} (a)), where one might assume that the controller's impact is minimal,
%}
%part of the wave after the AV has caught up with the bottleneck 
 %Within this region, 
 the speed variance in front of the AV is $\boldsymbol{1.73}\  m^2 \cdot s^{-2}$ while it reduces to $\boldsymbol{1.08}\  m^2 \cdot s^{-2}$ due to the AV's smoothing effect. 

\begin{figure}[h]
\centering
    \includegraphics[width=0.45\textwidth]{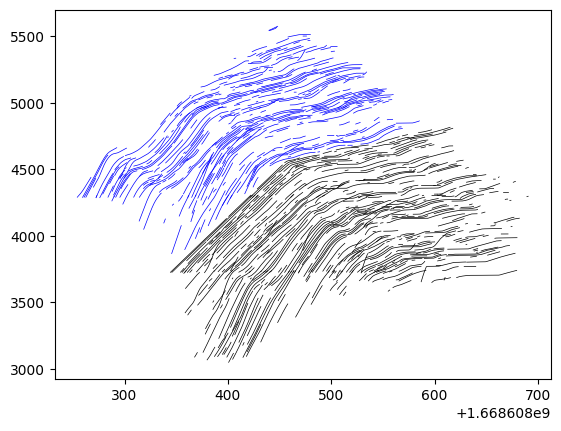}
    \caption{Time-space diagrams showing the trajectories of the vehicles up to 1500 m upstream (black) and up to 1400 m downstream (blue) of the AV. }
    \label{fig:distlong}
\end{figure}

\begin{figure}[h]
    \centering
    \includegraphics[width=0.45\textwidth]{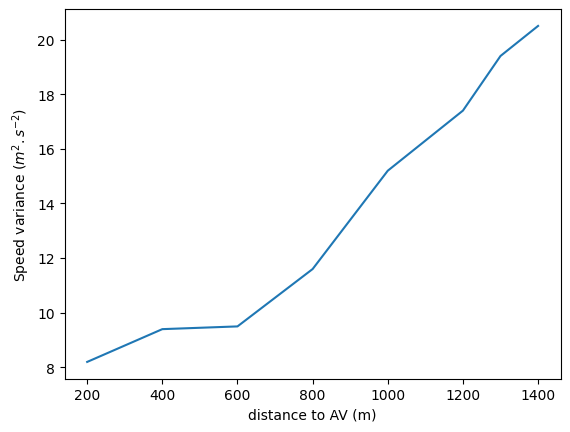}
    \caption{Speed variance of the vehicles behind the AV up to different distances.}
    \label{fig:disteffect}
\end{figure}

Our analysis thus far focuses on the vehicles within $400\ m$ behind the AV. It is worth noting that the effect of the AV diminishes with the distance. In Figure \ref{fig:distlong}, we extend the $400\ m$ to 
%$\textcolor{orange}{1400}\ m$ 
1400 m
to observe the behavior of the vehicles driving further behind the AV. We notice that speed oscillations start to reappear after around $600\ m$ behind the AV. We made this observation more concrete by reporting the speed variance of the vehicles behind the AV up to different distances ranging from $200\ m$ to $1400\ m$ in Figure \ref{fig:disteffect}. We notice a sharp increase in the speed variance after 600m. Furthermore, we report the percentage change in the speed variance considering different distances behind and in front of the AV in Table \ref{tab:metrics_sv}.\\
Finally, in Figures \ref{fig:pic1} and \ref{fig:pic2} we illustrate the physical vehicle and the research team that executed the experiment.

\section{Conclusion}

We presented an acceleration-based controller for a connected AV to smooth stop-and-go waves in highway traffic. The controller was implemented on a commercially available Toyota Rav 4, using measurements typically used by the Automated Cruise Control. The vehicle was deployed on the I-24 in the Nashville area in bulk traffic during rush hours, as part of a large-scale experiment. Trajectories were reconstructed thanks to the I-24 MOTION system using 276 high-fidelity cameras. The experimental results show the control ability to dampen stop-and-go waves in real traffic at peak hours. The controlled AV reduces the overall speed variance of the traffic 
up to $600m$ 
behind by around 50\%.
%\textcolor{red}{This is an incentive to XXX}

%%%%FOR LATER: EXPLAIN SENSITIVITY MPG%%%%
% Another limitation is the sensitivity of the metrics. Given that the measures are relatively local (between 200m-1km), they are subject to noise and particular behavior and therefore have to be taken with caution. 
% To illustrate this, in Figure \ref{fig:disteffect}, we show the speed variance of the vehicles behind the AV up to different distances from 200m to 1400m. We see a nearly strict increase, especially after 600m. Similarly In Table \ref{tab:metrics_sv} we take into consideration the speed variance in boxes of different size in front of the AV. While the dampening effect of the controller on the speed variance can be seen with all sizes of boxes, we see this local noise effect.

% \begin{figure}[h!]
%     \centering
% \includegraphics[width=0.95\linewidth]{figures/wave_lane_3_2000_800_800.png}\\ \includegraphics[width=0.95\linewidth]{figures/wave_1_2000_400_800.png}
%     \caption{Experimental time-space diagrams across a single macroscopic wave \textbf{(upper):} in the controlled AV's lane (lane 3) \textbf{(lower):} in a different lane (lane 1). \label{fig:time-space}}
% \end{figure}
%\textcolor{red}{Replace by a full figure without cutting the wave ?}

%\textcolor{red}{Limitations: Effect that starts to dissipate after a number of vehicles. Caveat on a reliable energy consumption}

\section{Acknowledgement}

This material is based upon work supported by the National Science Foundation under Grants CNS-1837244 (A. Bayen), CNS-1837652 (D. Work), CNS-1837481 (B. Piccoli), CNS-1837210 (G. Pappas), CNS-1446715 (B. Piccoli), CNS-2135579 (D. Work, A. Bayen, J. Sprinkle, J. Lee) as well as the IEA project SHYSTRA and the PEPS JCJC 2022 of CNRS INSMI (A. Hayat and S. Xiang). This material is based upon work supported by the U.S.\ Department of Energy's Office of Energy Efficiency and Renewable Energy (EERE) under the Vehicle Technologies Office award number CID DE--EE0008872. The views expressed herein do not necessarily represent the views of the U.S.\ Department of Energy or the United States Government.

\appendix
\section{Identifying moving border of the wave}
\label{app:cutting}
\begin{figure}
    \centering
    \includegraphics[width=0.45\textwidth]{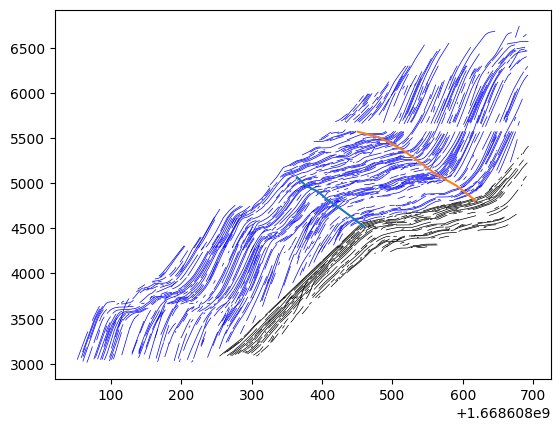}
    \caption{\textbf{Automatic identification of the congested part of the wave.} The frontiers of the congested part of the wave are represented in blue and orange. We display all the trajectories located within 1500m in front of the AV and 400m behind.}
    \label{fig:identification}
\end{figure}

To identify the moving border of the wave for the cars in front and behind the AV we proceed as follows:
\begin{enumerate}
\item Define how far in front of the AV and how far behind the AV to consider (for example, 1500m in front, 700m behind)

%\item Mesh the space in front and behind the AV and aggregate cars in space intervals from the AV.
%\item Linearly interpolate the trajectory of the AV within the region of interest
\item Discretize the space in front and behind the AV in \textbf{boxes} of same width following the contour of the AV trajectory and ordered by distance to the AV (for example box 1 contains all trajectories starting between 0m and 200m from the AV, box 2 all trajectories starting between 200m and 400m, etc.). 
%\
%\item
%\item Determine the number of space-axis boxes to split this region into (11, resulting in 200m width space-axis boxes)
%\item Discretize the space-axis, following the contour of the linearly-interpolated AV trajectory (resulting in 11, 200m width \textit{boxes} following the contour of the AV trajectory)
%\item For each \textit{box}, create a list containing all trajectories that begin in that \textit{box}

%\item For each \textit{box}, list all trajectories that begin in that \textit{box}
%\item Denote \textit{boxes-front} to be \textit{boxes} in front of the AV trajectory and \textit{boxes-behind} to be \textit{boxes} behind the AV trajectory

\item Define a time-axis length to discretize each \textbf{box} with time (for example 10s)

\item For each \textbf{box}, create \textit{sub-boxes} by using the time-axis discretization length. With a 10s time discretization and a 200m space discretization, this results in 10s by 200m \textit{sub-boxes} for each \textit{box} following the contour of the AV trajectory)

%\item for each \textit{sub-box} iterate over the trajectories within the \textit{box}, creating dictionaries that hold speeds and positions of the trajectories
\item For a given \textbf{box}, compute the average speed and position of the vehicles in each \textit{sub-box}. Optionally, 
%if the discretization is very fine, 
use a moving average to smooth these values within the \textbf{box}.

\item Define a \textit{speed threshold} which will determine when the congested part of the wave begins or ends (for example 4 meters per second)
\item For each \textit{box}, iterate over the \textit{sub-boxes}, and compare the average speed in the current \textit{sub-box} to the average speed of the previous \textit{sub-box} and the \textit{speed threshold}. If the current average speed is less than the \textit{speed threshold} and the previous average speed is greater than the \textit{speed threshold}, denote the current \textit{sub-box} as the beginning of the wave. If the current average speed is greater than the \textit{speed threshold} and the previous average speed is less than the \textit{speed threshold}, denote the current \textit{sub-box} as the end of the wave. %\textcolor{red}{This is illustrated in Figure \ref{}.}

\item For each \textit{box}, define the boundary of the wave using the average time and average position of the \textit{sub-boxes} denoting the start and end of the wave. Optionally smooth the obtained frontiers using a moving average. 
\end{enumerate}
With 200m and 10s space and time resolution, this results in the identification of the congested part of the wave represented in Figure \ref{fig:identification}

%\section{Speed-profile}
\begin{figure}
    \centering
\includegraphics[width=0.45\textwidth]{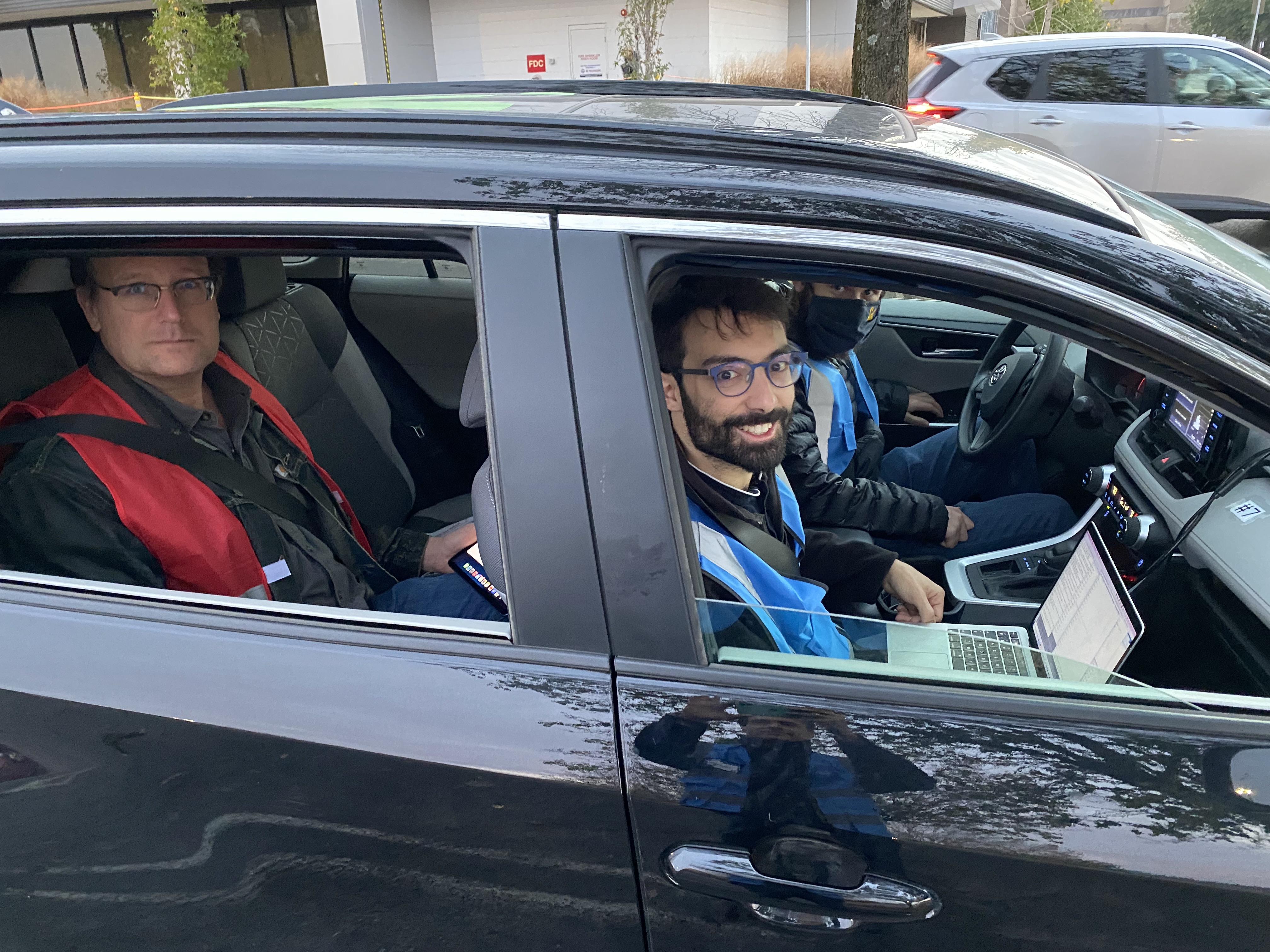}
    \caption{Test run on Wednesday November, 16th. The driver is assisted by a researcher visualizing and monitoring the control in real-time.}
    \label{fig:pic1}
\end{figure}

\begin{figure}
    \centering
\includegraphics[width=0.50\textwidth]{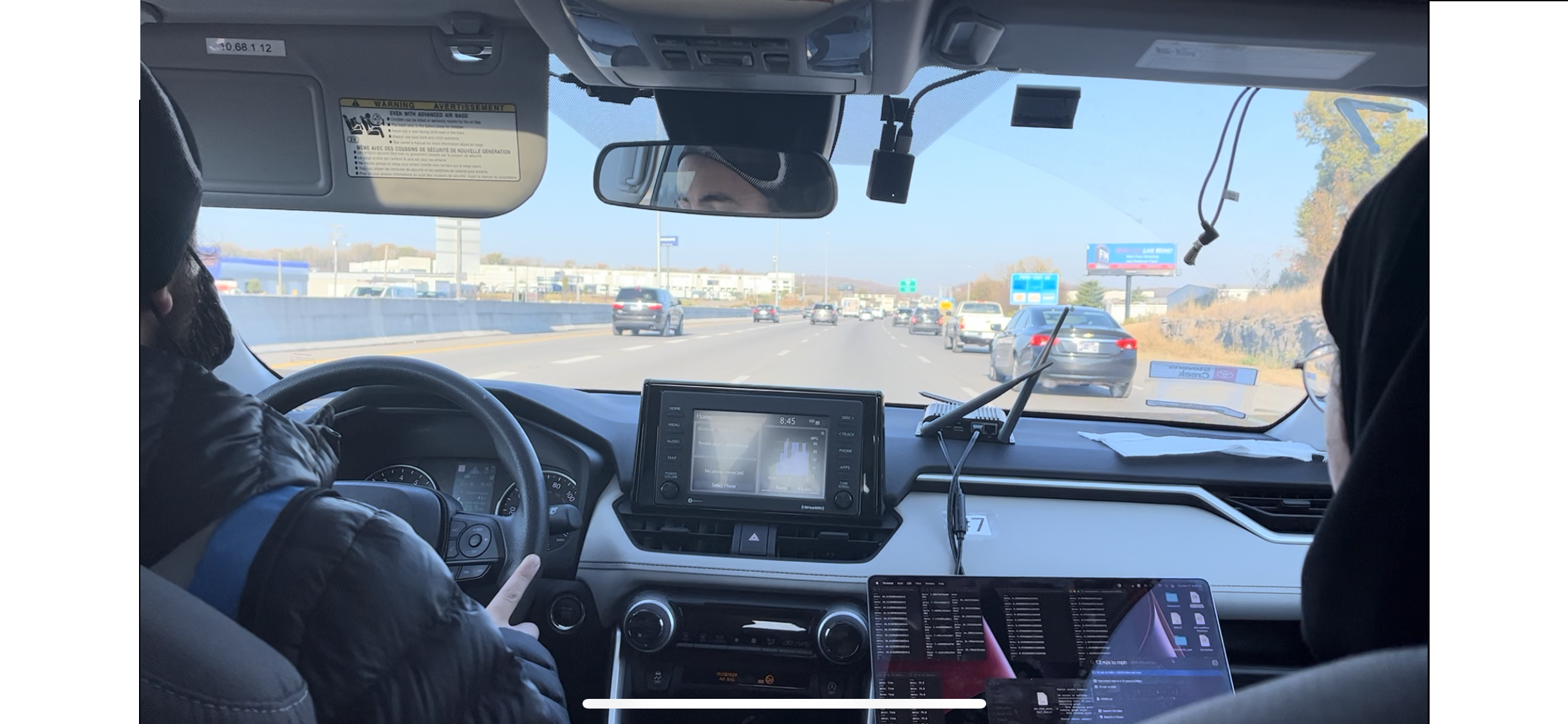}
    \caption{Follow-up run on Thursday November 17th. The driver activated the controller whose activity is monitored on a computer by the passenger}
    \label{fig:pic2}
\end{figure}

\bibliographystyle{plain}
\bibliography{references,CIRCLES_key_papers_arxiv}
%OK seen
%\endarticle

\end{document}